\documentclass[useAMS,usenatbib]{mn2e}
\usepackage{lscape,epsfig}
\usepackage{graphicx}
\usepackage[margin=5pt,font=small,labelfont=bf,
            justification=raggedright,singlelinecheck=false]{caption}
\title[Absolute properties of BG Ind]
 {Absolute properties of BG Ind -- a bright F3 system 
  just leaving the Main Sequence\thanks
   {Based in part on data obtained at the South African Astronomical
    Observatory.}
 }
\author [M. Rozyczka et al.]
 {M. Rozyczka,$^{1}$\thanks{E-mail: mnr@camk.edu.pl}
  J. Kaluzny,$^{1}$
  W. Pych$,^{1}$
  M. Konacki$,^{2}$
  \newauthor
  K. Ma\l ek,$^{3}\ddag$
  L. Mankiewicz,$^{3}\ddag$
  M. Soko\l owski,$^{4}\ddag$
  A.F. \.Zarnecki$^{5}$\thanks{for the Pi of the Sky team}
  \\
  $^{1}$Nicolaus Copernicus Astronomical Center, Bartycka 18, 00-716 Warszawa, Poland\\
  $^{2}$Nicolaus Copernicus Astronomical Center, Rabia\'nska 8, 87-100 Toru\'n, Poland\\
  $^{3}$Center for Theoretical Physics PAS, Al. Lotnik\'ow 32/46, 02-688 Warszawa, Poland\\
  $^{4}$The Andrzej So\l tan Institute for Nuclear Studies, Ho\.za 69, 00-681 Warszawa, Poland\\
  $^{5}$Institute for Experimental Physics, University of Warsaw, Ho\.za 69, 00-681 Warszawa, Poland
 }

\begin{document}

\date{Accepted ... Received ... in original form ...}

\pagerange{\pageref{firstpage}--\pageref{lastpage}} \pubyear{....}

\maketitle

\label{firstpage}

\begin{abstract}
We present photometric and spectroscopic analysis of the bright
detached eclipsing binary BG Ind. The masses of the components
are found to be $1.428\pm0.008$ and $1.293\pm0.008$ $M_\odot$ 
and the radii to be $2.290\pm0.017$ and $1.680\pm0.038$ $R_\odot$ 
for primary and secondary stars, respectively. Spectra- and 
isochrone-fitting coupled with colour indices calibration yield 
$[Fe/H]=-0.2\pm0.1$. At an age of 2.65$\pm$0.20 Gyr BG Ind is
well advanced in the main-sequence evolutionary phase - in fact, its
primary is at TAMS or just beyond it. Together with three similar
systems (BK Peg, BW Aqr and GX Gem) it offers an interesting
opportunity to test the theoretical description of overshooting in
the critical mass range 1.2 -- 1.5 $M_\odot$.
\end{abstract}

\begin{keywords} binaries: eclipsing -- stars: fundamental parameters --
stars: evolution
\end{keywords}

\section {Introduction}
 \label{sect: intro}

The 6th magnitude star BG Ind was classified as an F3 dwarf by
\citet{Mal75}, and several years later established as an eclipsing
binary by \citet{Man84} and \citet{Mat86}, who identified it as a 
detached system with partial eclipses and a period of 1.464047 d 
based on $ubvy$ data from 1984. The first radial velocity measurements
of BG Ind were performed by \citet{And84}. As they wrote, ``The
plates immediately showed spectral lines of two components of rather
similar type, the lines of one component (the primary) being
noticeably stronger and broader than those of the secondary.''
Although their data were rather scarce (just two spectra), the
resulting estimates of masses and radii of the components
($m_1\sim1.4$ $M_\odot$, $R_1\sim2.0$ $R_\odot$ and $m_2\sim1.2$
$M_\odot$, $R_2\sim1.5$ $R_\odot$) proved to be surprisingly
accurate. Additional Str\"omgren photometry was collected in 1986 by
\citet{VHM} who also solved the $vby$ light curves of the system,
and found an improved value of the period ($P = 1.464069$ d). For
the fixed mass ratio $q=0.85$ taken from Andersen et al. (1984) they
obtained $m_1=1.41$ $M_\odot$, $R_1=2.22$ $R_\odot$ and $m_2=1.20$
$M_\odot$, $R_2=1.60$ $R_\odot$.

Revised data of \citet{Mat86} and VHM, together with additional
points from 1987, were catalogued by \citet{Man91} and \citet{Ste93}
as a part of the {\it Long Term Photometry of Variables} program
conducted at ESO. BG Ind was monitored by Hipparcos satellite
\citet{Per97} and robotic telescopes ASAS \citep{Poj01} and Pi of
the Sky \citep{Mal10}; it is also included in {\it The
Geneva-Copenhagen survey of the solar neighborhood} \citep{Hol09}.
Recently, a spectroscopic solution of BG Ind based on data collected
in 2006 has been published by \citet{Bak10}, who for $P = 1.464069$
d and $i = 74.14^\circ$ found by VHM obtained $m_1=1.47\pm0.01$
$M_\odot$, $m_2=1.31\pm0.01$ $M_\odot$, and a semimajor axis
$a=7.64\pm0.04$ $R_\odot$.

With so much data available, BG Ind may seem to be well explored and
hence of little interest. However, there are at least three reasons
to investigate it more thoroughly than in the papers mentioned
above. First, VHM encountered problems with phasing, and they
recommended ``further monitoring of this binary system for minimum
times in order to obtain improved ephemeris, and to allow a study of
the behavior of its orbital period''. Second, systemic radial
velocity measurements gave diverging results: $v_\gamma = 39.8\pm4$
km s$^{-1}$ (VHM) and $59.4\pm5$ km s$^{-1}$ \citep{Bak10}. To make
the confusion even larger, from three velocity components of BG Ind
with respect to the Sun listed by \citet{Hol09} one obtains the {\it
total} velocity of $20.3$ km s$^{-1}$.

Third, and most important, both components of BG Ind belong to the
interesting mass range of 1.1--1.5 $M_\odot$ in which convective
cores begin to develop, affecting evolutionary tracks and isochrones
via overshooting-related uncertainties. Moreover, their masses are
remarkably similar to those of BK Peg -- a main sequence binary
recently studied by \citet{Cla10}. Since the metallicity is similar
in both cases, and smaller stellar radii observed in BK Peg
($R_1=1.987\pm0.008$ $R_\odot$, $R_2=1.473\pm0.017$ $R_\odot$)
indicate a slightly less advanced evolutionary stage, these two
systems are a potential source of valuable information concerning
the endphases of hydrogen burning in stars somewhat more massive
than the Sun. Bearing this in mind, we decided to re-analyze BG Ind
with the aim to determine its precise physical parameters and
evolutionary status.

Our paper is based on photometric and spectroscopic data  described
in Sect. 2. The analysis of the data is detailed in Sect. 3, and its
results are discussed in Sect. 4.
\section {Observational material and data reduction}
 \label{sect: Observations}
\subsection {Photometry}
\label{sect: obsphot}

Our photometric solutions are based on six sets of photometric measurements
listed in Table \ref{tab: obsphot}, spanning a period from May 1986
to November 2009 (points with $\Delta m > 0.1$ mag, where $\Delta m$
stands for the deviation from the preliminary fit (see Sect.
\ref{sect: tempsec}), were rejected from the original ASAS and Pi
datasets). We found that all photometric data from Table \ref{tab: obsphot}
neatly phase with the ephemeris
\begin{eqnarray}
 t_0 &=& {\rm HJD}~47876.3792~\pm{\rm0.0004}
 \label{eq: ephem}\\
 P   &=& 1.46406335~\pm0.00000002~{\rm d} \nonumber
\end{eqnarray}
(see Fig. \ref{fig: phasing}), and we are rather confident that the period of BG Ind remained
constant for over 23 years. The only indication suggesting a possible period
change comes from the earliest observations of this system (runs 1 and 2 of
VHM, not included in our analysis).
\setlength{\tabcolsep}{5pt}
\begin{table}
  \caption{List of light curves used in this paper for photometric solutions of BG Ind.
  \label{tab: obsphot}}
   \begin{tabular}{lccccc}
    \hline
     Label & Filter & 1$^{\mathrm st}$ day  & last day  &  Number of  & Ref.  \\
           &        & \multicolumn{2}{c}{[HJD-2400000]} &  data points&       \\
    \hline
     MSb  & Str\"omgren $b$   &  44582& 47070&  165 & 1,2\\
     MSv  & Str\"omgren $v$   &  46582& 47070&  172 & 1,2\\
     MSy  & Str\"omgren $y$   &  46582& 47070&  172 & 1,2\\
     HipH & Hipparcos $H$     &  47888& 49054&  142 & 3  \\
     ASV  & Johnson $V$       &  51997& 55167&  576 & 4  \\
     PiR  & Cousins R         &  53902& 54933&  828 & 5  \\
    \hline
   \end{tabular}\\
\rule{0 mm}{4 mm} {\small 1: \citet{Man91}; 2: \citet{Ste93}; 3:
\citet{Per97}; 4: \citet{Poj01}; 5: \citet{Mal10}. Pi of the Sky 
observations were done wihtout filters, but the combined sensitivity 
of optical system and detector closely matched that of Cousins R-filter.}
\end{table}

\noindent Available online are also Hipparcos $B_T$ and $V_T$
lightcurves \citep{Per97}, and Str\"omgren $u$ lightcurve of
\citet{Man91} and \citet{Ste93}. Their quality was too poor to use
them for photometric solutions; however $B_T$ and $V_T$ data yielded
a useful estimate of the temperature of the hotter component (see
Sect. \ref{sect: tempsec}). We note here that according to
\citep{Suc03} the Str\"omgren $(b-y)$ excess of BG Ind amounts to
0.001 only, so that reddening effects can be neglected.
\subsection {Spectroscopy}
\label{sect: obsspec} Because of problems with the systemic velocity
of BG Ind mentioned in Sect. \ref{sect: intro} we decided to use our
own spectral data, collected in September / October 2007 with the
fiber-fed Giraffe spectrograph on the 1.9-m Radcliffe telescope at
the South African Astronomical Observatory. The seeing oscillated
between 1.5 arcsec and 2.5 arcsec. A 2.7 arcsec entrance window
provided a resolution of almost $R = 40000$ at $\lambda = 4470$
\AA. During the observations pairs of scientific spectra were taken,
separated by an exposure of a thorium-argon hollow-cathode lamp. The
exposure times per spectrum ranged from 400 s to 900 s. The
observations were reduced within the IRAF \footnote {IRAF is
distributed by the National Optical Astronomy Observatories,
 which are operated by the AURA, Inc., under cooperative agreement
 with the NSF.}
ECHELLE package. After bias and flat-field correction each pair of
the frames was  combined into a single frame, allowing for the
rejection of cosmic ray hits. Altogether, 23 reduced spectra were
obtained. For further analysis a wavelength range extending from
4300{\AA} to 5800{\AA} was used, in which most of the spectra had
$20<\mathrm{S/N}<40$ with a few cases of lower quality.

Our method of radial velocity measurements is based on the
broadening  function formalism introduced by \citet{Ruc02} and it
was described in detail by \citet{Kal06}. \footnote {The software
used in this paper is available at
http://users.camk.edu.pl/pych/BF/.} The spectrum of the bright star
HD 200163 (spectral type F3V), obtained in the same observing period
and with the same instrumental setup, served as a template. The
resulting barycentric radial velocities are listed in Table
\ref{tab: radvel}. The mean error, estimated from velocity-curve
fitting, is $\pm$0.63 km s$^{-1}$
\begin{table}
  \caption{Barycentric radial velocities of BG Ind phased according
           to the ephemeris given by equation (\ref{eq: ephem}).
           \label{tab: radvel}}
   \begin{tabular}{ccrr}
    \hline
    HJD-2454000 & Phase  & $v_{\rm 1}$ [km s$^{-1}$]& $v_{\rm 2}$ [km s$^{-1}$]\\
    \hline
     363.398889 &0.833 &134.234  &-83.756\\
     363.420815 &0.848 &127.712  &-76.513\\
     366.290933 &0.808 &142.702  &-91.436\\
     368.315269 &0.191 &-78.691  &152.746\\
     368.335724 &0.205 &-83.175  &157.484\\
     368.365740 &0.225 &-87.252  &161.310\\
     369.244426 &0.825 &136.888  &-85.218\\
     369.263211 &0.838 &131.761  &-79.637\\
     371.249086 &0.195 &-80.656  &153.364\\
     371.266822 &0.207 &-82.224  &157.092\\
     371.284471 &0.219 &-85.197  &159.509\\
     371.315563 &0.240 &-86.278  &161.496\\
     371.328299 &0.249 &-86.007  &162.845\\
     371.344954 &0.260 &-86.568  &162.832\\
     371.361949 &0.272 &-85.781  &160.540\\
     371.379009 &0.283 &-83.505  &158.200\\
     371.395454 &0.295 &-82.164  &157.302\\
     371.412749 &0.306 &-79.292  &153.182\\
     372.255302 &0.882 &112.450  &-57.583\\
     372.275909 &0.896 &104.275  &-48.576\\
     372.292410 &0.907 & 96.605  &-41.796\\
     372.308893 &0.918 & 89.945  &-33.640\\
     376.386726 &0.221 &144.054  &-92.932\\
    \hline
   \end{tabular}
\end{table}

The binary spectrum of BG Ind was disentangled using the code of
\citet{Kon10} with the aim to estimate the temperature of the
components. Unfortunately, because of strong rotational broadening
and blending of the lines the only method we could apply was a
direct comparison to synthetic spectra (see Sect. \ref{sect:
tempsec}).
\section {Analysis}
 \label{sect: analysis}
The components of BG Ind have masses in the range 1.1--1.5 $M_\odot$
(VHM, \citeauthor{Bak10} \citeyear{Bak10}), so that their envelopes
should be convective. Consequently, we adopted gravity darkening
exponents $g1=g2=0.32$ and bolometric albedos $A1=A2=0.5$. Just in
case, we also obtained a solution with radiative envelopes
($g1=g2=A1=A2=1.0$) and we found that even within unrealistically
small formal error limits calculated by PHOEBE which are  all
``radiative'' parameters but one did not differ from the
``convective'' ones. The exception was the radius of the primary
$R_1$ which in the radiative case was by 5.6\% smaller (we define
primary as the {\it more massive} component).

Since BG Ind is a short-period system, and broadening-function
fitting yielded $v\sin i$ values close to those expected from the
synchronous rotation for the radii obtained by VHM, we assumed full
synchronization of both components. The effects of reflection were
included, and a logarithmic limb-darkening based on tables by
\cite{VHa93} was used as implemented in PHOEBE 031a.

\begin{figure}
\includegraphics[width=8 cm,bb= 38 370 565 692,clip]{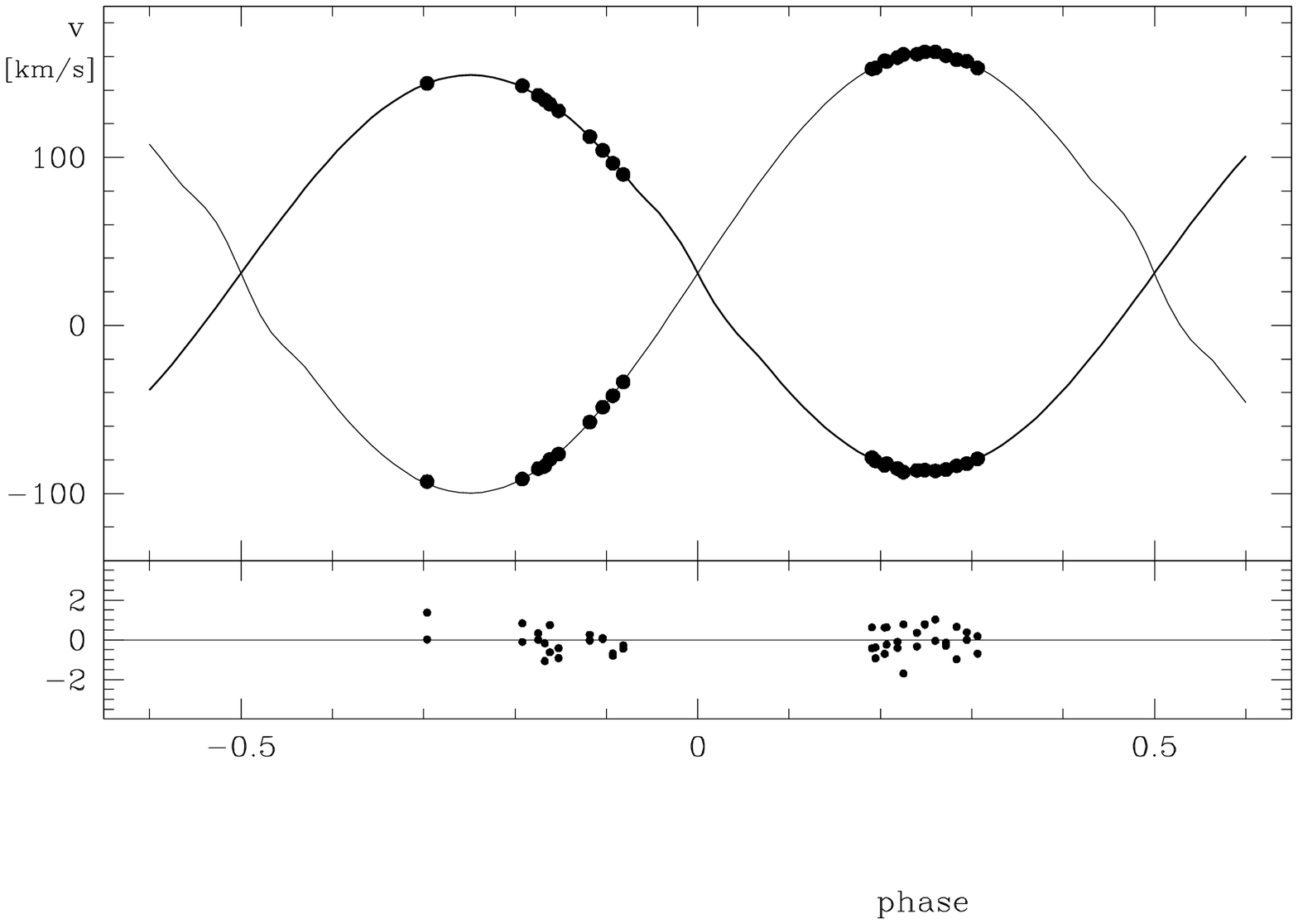}
 \caption {Velocity curve of BG Ind based on data contained in Table
           \ref{tab: radvel}. Systemic velocity and rms residuum are equal to
           $31.2 \pm 0.1$ km s$^{-1}$ and 0.63 km s$^{-1}$ respectively. Phase 0
           corresponds to the center of the {\it shallower} photometric minimum.
           \label{fig: radvel}
          }
\end{figure}

According to \cite{Hol09} the system is slightly underabundant in
metals ($[Fe/H]=-0.3$). However, their estimate is based on
Str\"omgren photometry of the {\it total} light, and as such it
cannot be 100\% reliable. As for the temperatures of the components,
VHM assumed $T_1 = 7000$ K for the primary, and obtained $T_2 =
6450$ K for the secondary from the fit. While these values are
certainly reasonable, they may be too high for the radii obtained by
VHM which indicate that at least the primary is about to leave the
main sequence or even has left it. Moreover, upon comparing our
phased velocity curve with phased light curves from Table \ref{tab:
obsphot} we found that the hotter component of BG Ind is {\it the
secondary} (see Fig. \ref{fig: radvel}). \footnote {The
misidentification of VHM is entirely understandable when one
remembers that they had just two velocity measurements and only an
approximate value of the period.} The same conclusion follows from
the velocities measured by \cite{Bak10} which also nicely phase with
the ephemeris (\ref{eq: ephem}), showing, however, a much larger
scatter (the rms residual from our fit to their data is 4.6 km
s$^{-1}$, only slightly larger than 4.4 km s$^{-1}$ found in the
original paper).

\noindent
Thus, while searching for an observational estimate of the
temperature of BG Ind components we decided to focus on the
secondary. As it rotates significantly slower than the primary and
must be less evolutionary advanced, we expected it to be a more or
less normal main-sequence star to which available color-temperature
calibrations could be reliably applied.
\subsection{Preliminary solutions}
We started the analysis from searching for solutions with $T_2$
equal to 7000, 6500 or 6000 K, and $[Fe/H]$ equal to 0.0 or -0.3. We
used PHOEBE interface \citep{Prs05} to the Wilson-Devinney code
\citep{WD71}, solving for all lightcurves simultaneously. The
essential aim of preliminary calculations was to check how strongly
in this parameter range vary the gravitational acceleration of the
secondary $g_2$ and the contributions of the secondary to the total
light in various wavebands at $\varphi=0.25$. As detailed in the
next Section, the first of those parameters was needed for an
estimate of $T_2$ based on our spectra, while the remaining ones --
for an independent estimate of $T_2$ based on calibrations of color
indexes.

There is no indication for a nonzero eccentricity in either light or
velocity curves, but just in case we included iterations of $e$.
The solutions yielded $e=0.0\pm0.0005$, and an almost constant
$\log g_2=4.14-4.15$. The primary's acceleration $g_1$ was also almost
constant, but smaller by a factor of 2, i.e. appropriate
for the beginning of the subgiant branch rather than for the main sequence.
\setlength{\tabcolsep}{8pt}
\begin{table*}
\begin{minipage}{126 mm}
  \caption{PHOEBE solutions with $T_2=6650$ K and $[Fe/H]=-0.2$. \label{tab: phoebesol}}
   \begin{tabular}{lrcccccccccc}
    \hline
 light & rms  &  $i$  & $a$  & $m_1$ & $m_2$ & $R_1$ & $R_2$ &  $T_1$ & $L_1$ & $L2$ \\
 curve &[mmag]&   [deg] & [$R_\odot$] & [$M_\odot$] & [$M_\odot$] & [$R_\odot$] &
 [$R_\odot$]  & [K]  & [$L_\odot$] &[$L_\odot$]\\
    \hline
 MSb   &  7.0 & 72.95 &  7.567 & 1.428 & 1.293 & 2.313 & 1.651 & 6312 & 7.37 & 4.62\\
 MSv   &  7.5 & 72.95 &  7.567 & 1.428 & 1.293 & 2.303 & 1.663 & 6320 & 7.34 & 4.69\\
 MSy   &  7.9 & 73.00 &  7.565 & 1.427 & 1.292 & 2.165 & 1.796 & 6443 & 7.00 & 5.47\\
 HipH  &  6.4 & 73.12 &  7.560 & 1.424 & 1.289 & 2.226 & 1.723 & 6382 & 7.13 & 5.04\\
 ASV   & 16.1 & 72.72 &  7.577 & 1.434 & 1.298 & 2.395 & 1.637 & 6375 & 8.22 & 4.55\\
 PiR   & 11.4 & 73.03 &  7.567 & 1.428 & 1.293 & 2.340 & 1.610 & 6262 & 7.30 & 4.39\\
    \hline
   \end{tabular}\\
\end{minipage}
\end{table*}

\subsection{The temperature of the secondary}
\label{sect: tempsec}
With $g_2$ fixed, we generated an array of synthetic spectra for
$6000 {\mathrm K} \le T \le 7000 {\mathrm K}$ and $-1 \le [Fe/H] \le
0$ based on the library of \citep{Coe05}. The respective spacings
$\Delta T$ and $\Delta [Fe/H]$ were equal to 100 K and 0.1. All
spectra were rotationally broadened with $v\sin i$ = 78 km s$^{-1}$
for the primary and 53 km s$^{-1}$ for the secondary, consistently
with the mean values of $i$ and component radii obtained from
preliminary fits (73$^\circ$, 2.35 $R_\odot$ and 1.61 $R_\odot$,
respectively). Next, for each synthetic spectrum the sum of squared
deviations from its observed counterpart was calculated. In the case
of the secondary the smallest sum was obtained for $T_2$ = 6500 K
and $[Fe/H]$ = -0.2 (see Fig. \ref{fig: sec_sp_comp} for the
comparison of best-fitting spectrum to the observed one). In the
case of the primary no unique minimum was found, for what broad
and/or blended lines are probably to be blamed (e.g. MgI lines at
5167{\AA} and 5173{\AA} were observed as a single spectral feature).
We estimate the uncertainties of temperature and metallicity at
$\pm100$ K and $\pm 0.1$, respectively.

In all preliminary solutions with $[Fe/H]=0$ the secondary to
primary luminosity ratio $L_2/L_1$ (found from the contribution of
each component to the total light of the system at $\varphi=0.25$)
was almost constant in each of $b$, $v$ and $y$ bands, amounting
respectively to 0.570-0.573, 0.598-0.606 and 0.551-553. Given the
total apparent magnitudes of BG Ind at $\varphi=0.25$, we used these
ratios to calculate the corresponding apparent magnitudes of the
secondary, and we found its temperature from $(b-y)-T_{eff}$
calibration of \cite{Cas10}. The result for $[Fe/H]=0.0$ was a
remarkably constant $T_2 = 6523-6528$ K. Solutions with
$[Fe/H]=-0.3$ were only slightly more diverging, with $T_2 =
6477-6487$ K. After averaging, we got $T_2 = 6525\pm63$ K and $T_2 =
6483\pm63$ K, respectively, where the error includes inaccuracies of
color-temperature calibration and magnitude measurement at
$\varphi=0.25$. Based on the calibrations of \cite{Hol07}, we also
derived the observed $[Fe/H]$ index, which varied between -0.4 and
-0.7. While being only marginally consistent with the lower of the
$[Fe/H]$ values assumed in preliminary fitting, it confirmed that BG
Ind is indeed metal-deficient.

The third estimate of $T_2$ was based on the Hipparcos photometry.
Proceeding as before, we found the light ratio $L_2/L_1$ at
$\varphi=0.25$ in $B_T$ and $V_T$ bands, and, using
$(B-V)_T-T_{eff}$ calibration of \cite{Cas10}, we derived $T_2 =
6650\pm83$ K for $[Fe/H]=0$ and $T_2 = 6496\pm83$ K for
$[Fe/H]=-0.3$.

For the fourth temperature estimate we used Hipparcos parallax of BG
Ind ($\pi = 15.04 \pm1.1$ mas) and ASAS photometry in $V$-band.
Proceeding as before and using bolometric corrections of
\cite{Lej95}, we found the observed $V-$band magnitude of the
secondary ($V_2^{o}$=7.22), and compared it to apparent magnitudes
$V_2^{c}$ calculated from preliminary solutions. Solutions with
$T_2=6000$ which yielded $7.59\le V_2^{c}\le7.88$ were clearly ruled
out. Those with $T_2=6500$ K and $T_2=7000$ K produced,
respectively, $7.27\le V_2^{c}\le7.56$ and $6.95\le V_2^{c}\le7.24$,
i.e. magnitudes almost consistent with $V_2^{o}$, but remaining on
the low side in the first case and on the high side in the second
case. A nearly ideal agreement between observed and calculated
magnitude was reached in an additional preliminary fit with $T_2 =
6750$ K, for which the difference $V_2^{o} - V_2^{c}$ was equal to
just 0.02 mag.

Thus, the revised Hipparcos parallax pulled the temperature estimate
firmly upward. Trusting it, we adopted $T_2 = 6650$ K with a
fiducial uncertainty range of $\pm100$ K, marginally compatible with
all available data. As it was detailed above, the trend observed in
preliminary solutions indicated that the metallicities obtained from
Str\"omgren photometry would cause $T_2$ to fall too low to be
compatible with $V_2^o$. Bearing this in mind, we set $[Fe/H]$ to
-0.2, i.e. to the value resulting from the spectroscopic estimate.
\subsection{Final model fitting}
\label{sect: modfit}
For the final analysis the JKTEBOP code \citep{Etz81, Pop81, Sou04a,
Sou04b} and a spectroscopic data solver written and kindly provided
by G. Torres were used except PHOEBE. While JKTEBOP gives realistic
estimates of the errors, it cannot be used when the stars are too
distorted. Unfortunately, this is the case of BG Ind, where the
primary's distortion exceeds the allowable limit by 10\%.
Consequently, we employed JKTEBOP solely to estimate the errors of
the photometric solution (note that differences between the errors
of the slightly inaccurate JKTEBOP solution and the errors of the
appropriate solution must be small quantities of the second order
which can be neglected). Similarly, realistic errors of the
spectroscopic solution were found using the Torres code. (Because
the Torres code requires center of mass velocities on input, it was
necessary to correct the observed light center velocities for
effects caused by the distortion of the components. Additive
phase-dependent corrections were calculated using the Wilson-Deviney
code; their values ranged from -0.86 to 0.27 km s$^{-1}$).
\begin{figure*}
\includegraphics[width=\textwidth,bb= 50 490 582 692,clip]{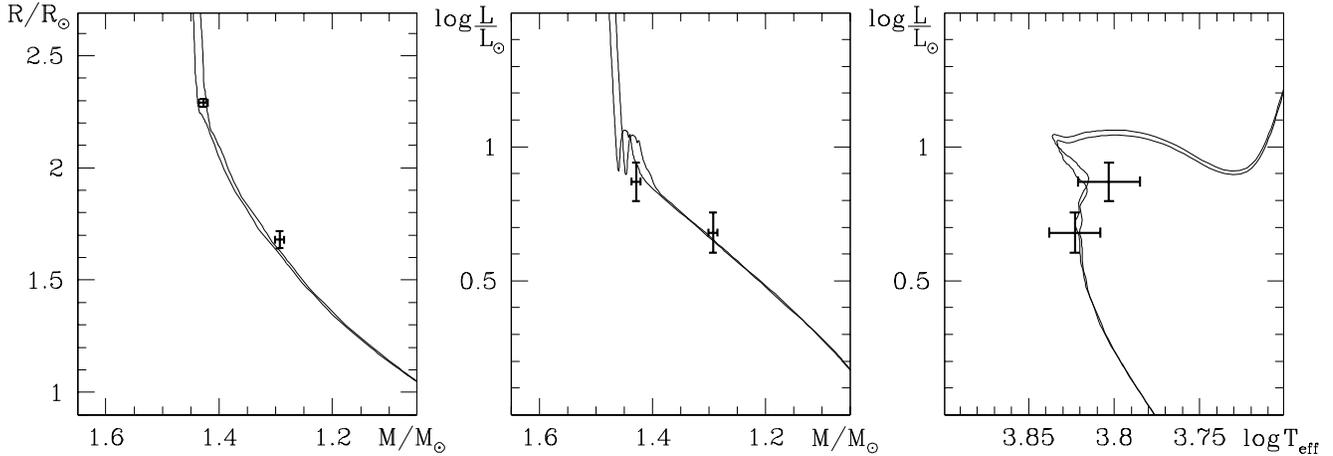}
 \caption {Location of BG Ind components on $M-R$, $M-\log L$ and
  $\log T_{eff} - \log L$ planes. The lines are Dartmouth $[Fe/H]=-0.2$
  isochrones for 2.60 and 2.67 Gyr (i.e. for extreme ages still fitting
  the primary on the $M-R$ plane).
 \label{fig: BGInd3}}
\end{figure*}

JKTEBOP cannot deal with multiple lightcurves, so that the final
model fitting had to be performed separately for each of the
photometric datasets listed in Table \ref{tab: obsphot}. We fixed
$e=0$ as indicated by preliminary solutions, and started the final
calculations from PHOEBE fits which produced six sets of parameters
listed in Table \ref{tab: phoebesol} whose second column shows the
rms deviation of observed points from the fit. Next, we calculated
errors of inclination $i$ and relative radii of the components
$r_{1,2}=R_{1,2}/A$ using JKTEBOP, and errors of $m_{1,2}\sin^3i$
and $A\sin i$ using the Torres code. The erros were then transformed
into errors of parameters returned by PHOEBE, and assigned to
respective PHOEBE solutions. Finally, weighted averages of PHOEBE
parameters and the errors of those averages were found from the
standard formulae
 \begin{equation}
 <x> = \frac{\sum_{j=1}^6 x_j/\sigma_j^2}{\sum_{j=1}^6 1/\sigma_j^2},
 \end{equation}
and
 \begin{equation}
 \sigma^2 = \frac{1}{\sum_{j=1}^6 1/\sigma_j^2}.
 \end{equation}
The final parameters with their errors are listed in Table
 \ref{tab: finpar}.

\section {Discussion and conclusions}
 \label{sect: Discussion}

In Sect. \ref{sect: intro} we outlined three problems which prompted
us to analyze BG Ind: variability of the period, doubtful systemic
velocity and lack of accurate parameters of the system. As mentioned
in Sect. \ref{sect: obsphot}, we found that between 1986 and 2009
the period remained constant at $1.46406335\pm0.00000002$ d. The
only indication that it might have changed comes from the earliest
observations of BG IND from 1981 and 1984 (runs 1 and 2 of VHM)
which were not included in our data. As for the systemic velocity,
our value of $31.2\pm0.2$ km s$^{-1}$ does not agree with any of
those obtained by other authors. The value found by VHM ($39.8\pm4$
km s$^{-1}$) was based on two measurements only, and its error was
likely underestimated. $59.4\pm5$ km s$^{-1}$ of \cite{Bak10} is
rather large for an F-type star in Sun's vicinity. The origin of
such a large discrepancy with all remaining estimates is difficult
to explain - we may only note that they did not observe any radial
velocity standards, and it is conceivable that they reversed the
sign of the  heliocentric correction while reducing the data.
Finally, the low $v_{total}=20.3$ km s$^{-1}$ of \cite{Hol09} can be
explained by the fact that they did not detect the binarity of BG
Ind (this system is not included in their Table 2), and their
velocity must have been contaminated by the orbital motion.
\begin{table}
  \caption{Physical parameters of BG Ind. \label{tab: finpar}}
   \begin{tabular}{lrcll}
    \hline
    Parameter & \multicolumn{3}{c}{Value} &Unit \\
    \hline
    $P$     &1.46406335&$\pm$& 0.00000002& [d]\\
    $i$     &  72.96   &$\pm$& 0.23  & [deg]     \\
    $e$     &  0.000     &$\pm$& 0.0005    &  \\
    $a$     &  7.567   &$\pm$& 0.013 & $R_\odot$ \\
    $m_1$   &  1.428   &$\pm$& 0.008 & $M_\odot$ \\
    $m_2$   &  1.293   &$\pm$& 0.008 & $M_\odot$ \\
    $R_1$   &  2.290   &$\pm$& 0.017 & $R_\odot$ \\
    $R_2$   &  1.680   &$\pm$& 0.038 & $R_\odot$ \\
    $\log T_1$[K] & 3.803 &$\pm$& 0.018 &      \\
    $\log T_2$[K] & 3.823 &$\pm$& 0.015 &      \\
    $\log L_1/L_\odot$ &   0.87 &$\pm$& 0.07 &  \\
    $\log L_2/L_\odot$ &   0.68 &$\pm$& 0.08 &  \\
    $[Fe/H]$ &-0.2    &$\pm$&0.1& \\
    age     &      2.65&$\pm$&0.20& Gyr\\
    \hline
   \end{tabular}
\end{table}

The parameters found in Sect. \ref{sect: modfit} are accurate enough
for isochrone fitting. We used solar-scaled Dartmouth isochrones
(Dotter et al. 2008) which include core overshooting defined as a
product of the pressure scale height and a factor $\alpha_{over}$
which depends on stellar mass and composition (at nearly-solar
metallicities it grows from 0.05 for $1.2\le M\le1.3 M_\odot$
through 0.1 for $1.3<M\le1.4 M_\odot$ to 0.2 for $M>1.4 M_\odot$).
The convection itself is treated according to the standard mixing
length theory with solar-calibrated mixing length parameter
$\alpha_{ml}=1.938$.
\begin{figure*}
\includegraphics[width=\textwidth,bb= 50 187 584 692,clip]{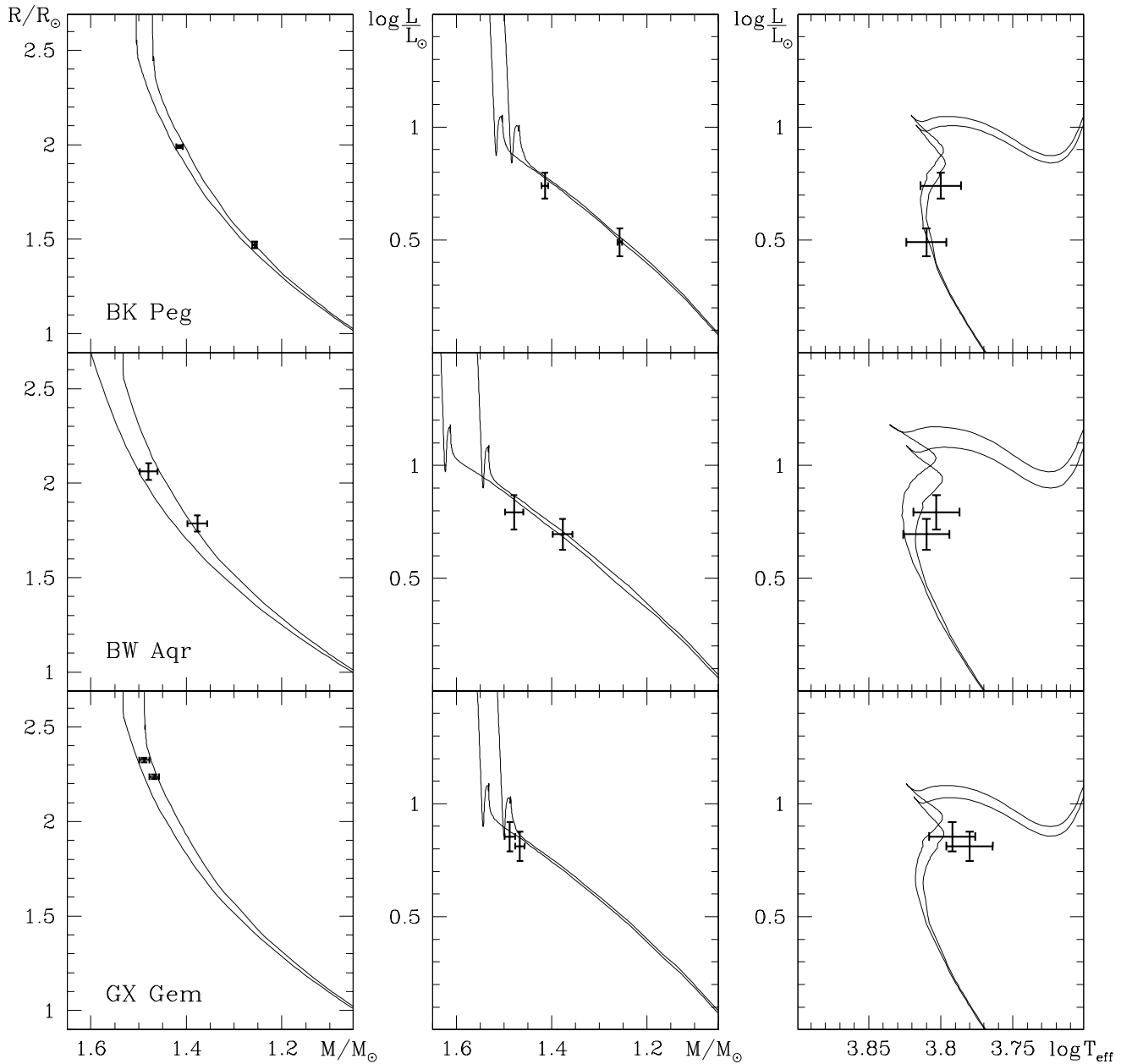}
 \caption {Systems with similar evolutionary advancement as BG Ind.
  The lines are Dartmouth $[Fe/H]=0.0$
  isochrones for 2.60 and 2.80 Gyr (BK Peg); 2.10 and 2.45 Gyr (BW Aqr);
  2.45 and 2.70 Gyr (GX Gem).
 \label{fig: othersystems}}
\end{figure*}

What makes BG Ind particularly interesting is that the masses of the
components fall at the beginning and at the end of the range where
$\alpha_{over}$ ramps up. First, we fitted isochrones calculated for
$-0.4\le[Fe/H]\le0$ with a step of 0.1 to the most accurately
determined parameters, i.e. masses and radii of the components. The
respective ages we found were 2.27-2.43, 2.45-2.55, 2.60-2.67,
2.77-2.85, and 2.96-3.10 Gyr. Next, we checked how well these
isochrones perform on $M-\log L$ and $\log T_{eff} - \log L$ planes.
The best agreement was obtained for $[Fe/H]=-0.2$. The fit with
$[Fe/H]=-0.1$ was almost equally good; that with $[Fe/H]=-0.3$ was
still acceptable; and those with $[Fe/H]=0 $ and $[Fe/H]=-0.4$ had
to be rejected. Thus, the isochrone fitting confirmed our
spectroscopic estimate of metallicity which we finally fixed at
$[Fe/H]=-0.2\pm0.1.$ The corresponding age is $2.65\pm0.20$ Gyr.

The location of BG Ind on $M-R$, $M-\log L$ and $\log T_{eff} - \log
L$ planes is shown in Fig. \ref{fig: BGInd3} together with $t =
2.60$ Gyr and $t = 2.67$ isochrones obtained for $[Fe/H]=-0.2$. One
can see that the more massive primary has almost reached the
beginning of the subgiant branch, while the secondary is still on
its way to TAMS. The agreement between theoretical and observational
data would be ideal if it were not for small discrepancies in $R_2$
and $T_1$ (by $\sim$3\% and $\sim$1.5\%, respectively). The first
one could originate from the fact that the eclipses of BG Ind are
partial and the secondary is by almost 40\% smaller than the primary
(and therefore less deformed). As a result, and because of rather
poor quality of available photometric data, the accuracy of $R_2$
determination has to be markedly lower than that of $R_1$. The
second discrepancy, at a first glance rather insignificant, turned
our attention because it occurs precisely where the effects of
overshoot-treatment should be largest (the primary of BG Ind is a
star with $M> 1.4 M_\odot$ at TAMS). We decided to check if the same
effect appears in other systems with similar masses and in similar
evolutionary phase.

Based on a recent compilation of \cite{Cla10}, we chose BW Aqr, BK
Peg, and GX Gem whose component masses range from 1.26 to 1.49
$M_\odot$, and whose $[Fe/H]$ indexes are consistent with 0. The
fitting of solar-scaled Dartmouth isochrones for $[Fe/H]=0$ yielded
respective ages of 2.15 - 2.45, 2.6 - 2.8 and 2.50 - 2.75 Gyr,
placed roughly halfway between Yonsei-Yale and VRSS ages quoted in
Table 12 of \cite{Cla10}. Figs. \ref{fig: BGInd3} and
 \ref{fig: othersystems} demonstrate that the temperature
discrepancy, absent in the relatively unevolved secondary components
of BG Ind and BK Peg, increases with the evolutionary advancement
(whose best indicator is the distance from the sharp upturn of the
isochrones on the $M-\log L$ plane) until it becomes clearly
visible in GX Gem whose both components are at TAMS or have already
left the main sequence. Note that luminosity errors are larger than
those quoted by \cite{Cla10} - this is because we recalculated them
according to the formula
 \begin{equation}
 \delta \log L = \sqrt{\left(2\frac{\delta R}{R}\right)^2 +
                      \left(4\frac{\delta T}{T}\right)^2}
 \label{eq: errL}
 \end{equation}
to make them consistent with ours.

The discrepancy is a 1-$\sigma$ effect and as such it may not be
real, however the trend it exhibits suggests there might be some
physics behind. It is beyond the scope of our paper to identify
physical factors or assumptions potentially responsible for such
effect. Whether anybody decides to look for them or not, improving
the quality of photometric and spectroscopic solutions of all four
systems is certainly a worthwhile task, although in the case of BG
Ind it might prove rather difficult because of strong rotational
broadening.

\section*{Acknowledgments}
We thank Guillermo Torres for providing the spectroscopic data solver.
Research of JK is supported by Foundation for Polish Science through 
the grant MISTRZ, and Polish Ministry of Science and Higher Education 
(PMSHE) through the grant N N203 379936 which also supports WP and MR. 
MK acknowledges the support from Foundation for Polish Science
through the FOCUS grant and fellowship, and European Research
Council through the Starting Independent Researcher Grant. Pi of the
Sky is financed by PMSHE in 2009-2011 as a research project.

\newpage
\setcounter{section}{1}
\setcounter{figure}{0}
\renewcommand{\thefigure}{\Alph{section}\arabic{figure}}
\begin{figure}
\section*{~}
\section*{appendix}
\section*{~}
\includegraphics[width=8cm,bb= 36 340 565
692,clip]{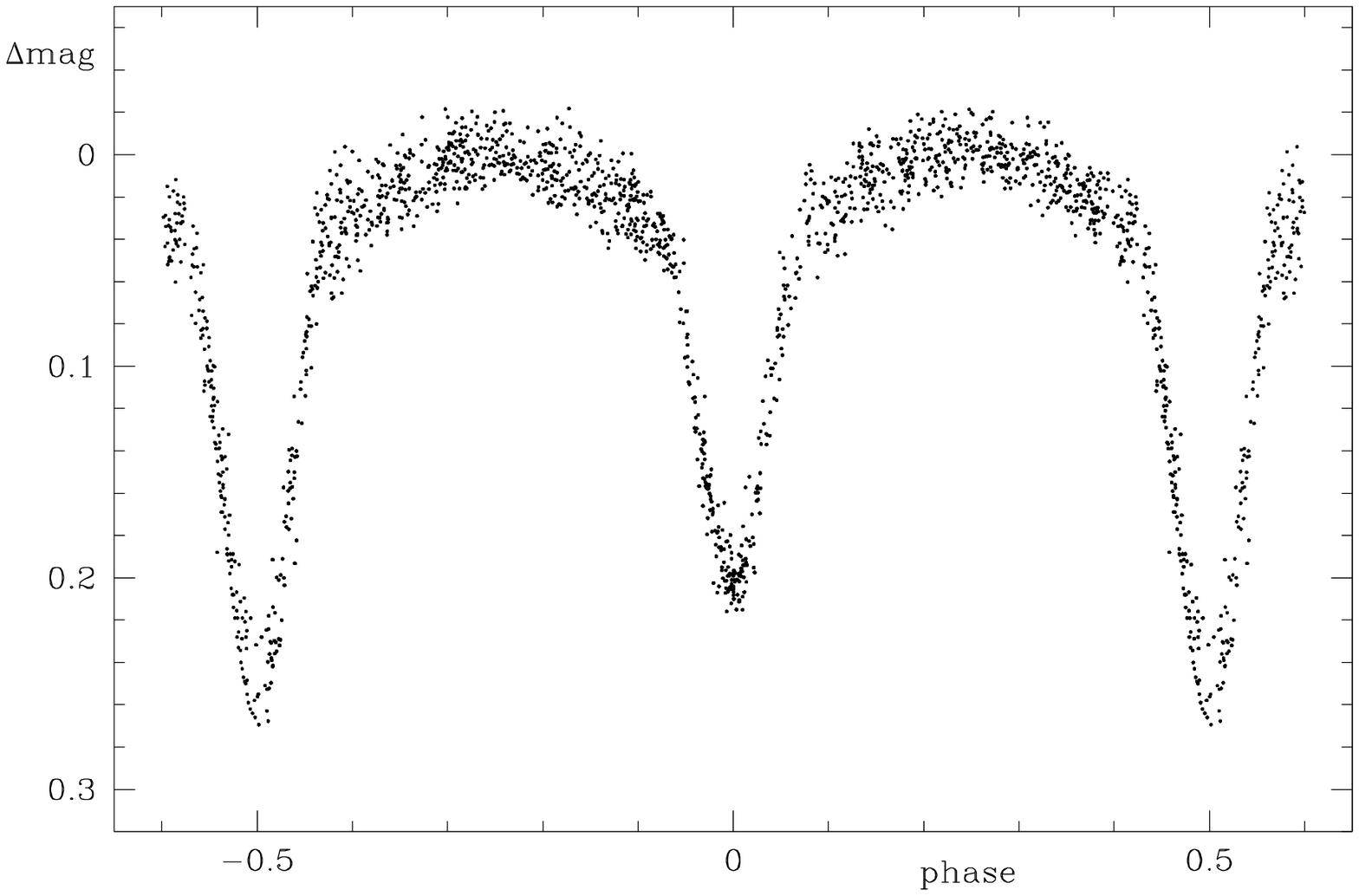}
 \caption {BG Ind light curves used in this paper (see Table \ref{tab: obsphot}
  for the list), phased with the ephemeris (\ref{eq: ephem}). Individual
  curves are normalized to magnitude 0 at maximum light.
 \label{fig: phasing}}
\end{figure}

\begin{figure}
\includegraphics[width=8cm,bb= 48 364 565 692,clip]{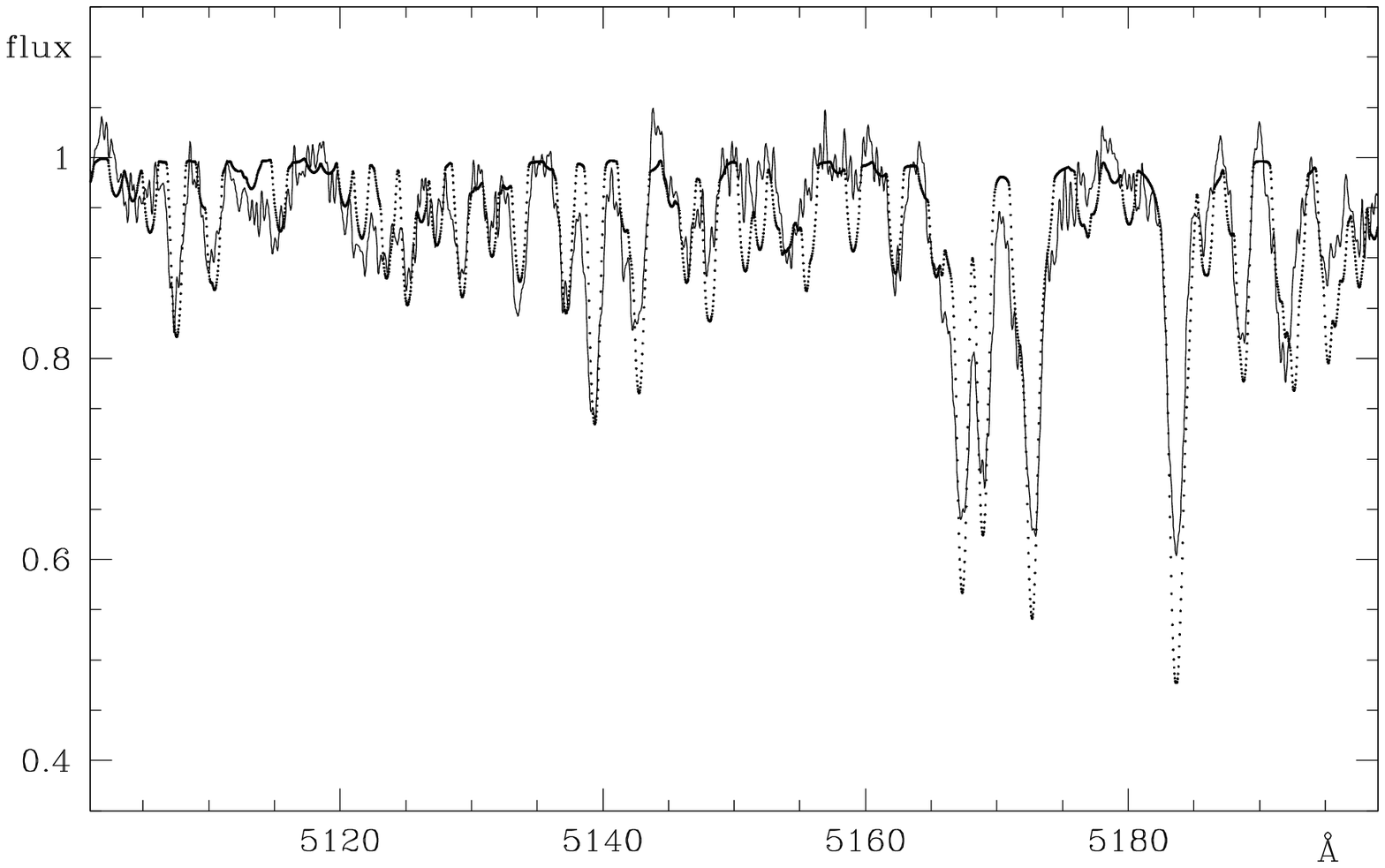}
 \caption {A section of the disentangled spectrum of the secondary (line)
  compared with the best fitting synthetic spectrum, obtained for $T_{eff}
  = 6500$ K, $g = 4.15$, [Fe/H] = -0.2, and rotationally broadened with
  $v\sin i$ = 53 km s$^{-1}$  (dots). The largest differences appear in deep
  lines and in overlap regions of echelle orders.
 \label{fig: sec_sp_comp}}
\end{figure}

\label{lastpage}

\end{document}